\documentclass[aps,prb,twocolumn,amsmath,amssymb,showpacs]{revtex4}
\usepackage{graphicx}
\usepackage{dcolumn}
\usepackage{bm}
\usepackage{color}
\usepackage{here}
\graphicspath{{fig/}}
\begin{document}

\title{Identification of an RVB liquid phase in a quantum dimer model with competing 
kinetic terms}

\author{Fran\c{c}ois Vernay$^1$, Arnaud Ralko$^1$, Federico Becca$^2$, and Fr\'ed\'eric Mila$^1$}

\affiliation{
$^1$Institute of Theoretical Physics, Ecole Polytechnique 
F\'ed\'erale de Lausanne, CH-1015 Lausanne, Switzerland \\
$^2$CNR-INFM Democritos National Simulation Centre and International School for 
Advanced Studies (SISSA), Via Beirut 2-4, I-34014 Trieste, Italy}

\date{June, 8th 2006}

\begin{abstract}
Starting from the mean-field solution of a spin-orbital model of LiNiO$_2$, 
we derive an effective quantum dimer model (QDM) that lives on the triangular lattice 
and contains kinetic terms acting on 4-site plaquettes and 6-site loops. 
Using exact diagonalizations and Green's function Monte Carlo simulations, we show that the 
competition between these kinetic terms leads to a resonating valence bond (RVB) 
state for a finite range of parameters. We also show that this RVB phase is connected to 
the RVB phase identified in the Rokhsar-Kivelson model on the same lattice in the context of
a generalized model that contains both the 6-site loops and a nearest-neighbor dimer 
repulsion. These results suggest that the occurrence of an RVB phase is a generic feature 
of QDM with competing interactions.
\end{abstract}

\pacs{75.10.Jm, 05.50.+q, 05.30.-d}

\maketitle

\section{Introduction}
After their first derivation by Rokhsar and Kivelson in 1988 in 
the context of cuprates,~\cite{rokhsar} the hard-core quantum dimer models (QDM)
have attracted significant attention. The phase diagrams of the QDM on the square 
and triangular lattices
have been investigated in great details,~\cite{leung,syljuasen} and, following the pioneering
work of Moessner and Sondhi on the triangular lattice,~\cite{moessner} the very existence
of a stable resonating valence bond (RVB) phase has been unambiguously 
demonstrated.~\cite{ralko}
The presence of a liquid phase with deconfined vison excitations~\cite{balents}
has also been established for a toy model living on the Kagome lattice.~\cite{misguich}

However, the relationship between QDMs and Mott insulators, the physical systems
for which they were
proposed in the first place, is not straightforward. It is well established
by now that the ground state of the S=1/2 Heisenberg model on the square and triangular
lattices exhibits long-range magnetic order of N\'eel and 120 degree type respectively,
and this type of order cannot be reached within the variational basis of Rokhsar and
Kivelson, which consists of short-range singlet dimers.
For a QDM to be a good effective model, one should thus identify models
for which the subspace of short-range dimer coverings on a certain lattice
is a good variational basis. 

The first example of such a case was provided by the S=1/2 Heisenberg model on the
trimerized Kagome lattice.~\cite{mila} Indeed, an effective spin-chirality model
living on a triangular lattice can be derived, and, at the level of a mean-field
decoupling between spin and chirality, the ground state manifold consists of all
dimer coverings on the triangular lattice. Going beyond mean-field is
thus expected to lead to a relevant effective QDM. Using Rokhsar and Kivelson's 
prescription, which consists in truncating the Hamiltonian and inverting the overlap 
matrix within the basis of dimer coverings, Zhitomirsky has derived such an
effective Hamiltonian~\cite{zhitomirsky} and shown that the main competition is between
kinetic terms involving loops of length 4 and 6 respectively, and {\it not} a competition
between a kinetic and a potential term, as for the QDM derived by Rokhsar and Kivelson. 
The next logical step would be to study the properties of this effective QDM. This is far 
from easy however. We know from the experience with the standard QDM model on the triangular 
lattice that the clusters reachable with exact diagonalizations are much too small to allow 
any significant conclusion regarding the presence of an RVB phase, and since there is no 
convention leading only to negative off-diagonal matrix elements, it is impossible to 
perform quantum Monte Carlo simulations.

Recently, two of us came across another model, for which the low
energy sector consists of almost degenerate singlet coverings on the triangular lattice.
This model is a Kugel-Khomskii~\cite{kugel} model that was derived in the 
context of LiNiO$_2$, and the mean-field equations that describe the decoupling of the spin 
and orbital degrees of freedom possess an infinite number of locally stable solutions. 
These solutions
are almost degenerate and correspond to spin singlet (and orbital triplet) dimers on the  
triangular lattice.~\cite{vernay} Following Rokhsar and Kivelson's prescription, an effective
QDM can also be derived (see Appendix~\ref{derivation}). As for the S=1/2 Heisenberg model 
on the trimerized Kagome lattice, it
consists of a competition between kinetic terms, with two important differences however.
The main term of length 6 lives on loops that have a shape of large triangles, a term absent 
in the other case. But more importantly, the off-diagonal matrix elements are {\it all} 
negative. 

Since the competition between kinetic processes was never investigated before, we have 
decided to concentrate on the minimal model obtained by keeping only the dominant term of 
length 6 for clarity. We have checked that the properties of the complete effective model 
are similar. This minimal model is described by the Hamiltonian:
\begin{widetext}
\begin{equation}\label{hamilt}
\begin{array}{rcl}
{ H}&=& \ -t \sum 
\left(
|\unitlength=1mm
\begin{picture}(6.2,5)
\linethickness{2mm}
\put(0.9,-.7){\line(1,2){1.8}}
\put(3.8,-.7){\line(1,2){1.8}}
\end{picture}
\rangle
\langle
\unitlength=1mm
\begin{picture}(6.5,5)
\linethickness{0.3mm}
\put(3.2,2.6){\line(1,0){3.2}}
\put(0.9,-.7){\line(1,0){3.2}}
\end{picture}
|
+h.c.\right)
- \ t^\prime \sum 
\left(
\left|\unitlength=1mm
\begin{picture}(7,6)
\linethickness{2mm}
\put(0.8,-1.7){\line(1,2){1.8}}
\put(6.4,1.6){\line(-1,2){1.8}}
\linethickness{0.2mm}
\put(3.8,-1.7){\line(1,0){3.6}}
\end{picture}
\right\rangle
\left\langle
\unitlength=1mm
\begin{picture}(7,6)
\linethickness{2mm}
\put(1.8,1.6){\line(1,2){1.8}}
\put(7,-1.7){\line(-1,2){1.8}}
\linethickness{0.2mm}
\put(-0.6,-1.7){\line(1,0){3.6}}
\end{picture}
\right|
+h.c.\right)
+ \ V \sum \left(
|\unitlength=1mm
\begin{picture}(6.2,5)
\linethickness{2mm}
\put(0.9,-.7){\line(1,2){1.8}}
\put(3.8,-.7){\line(1,2){1.8}}
\end{picture}
\rangle
\langle
\unitlength=1mm
\begin{picture}(6.2,5)
\linethickness{2mm}
\put(0.9,-.7){\line(1,2){1.8}}
\put(3.8,-.7){\line(1,2){1.8}}
\end{picture}|+
|
\unitlength=1mm
\begin{picture}(6.5,5)
\linethickness{0.3mm}
\put(3.2,2.6){\line(1,0){3.2}}
\put(0.9,-.7){\line(1,0){3.2}}
\end{picture}\rangle
\langle
\begin{picture}(6.5,5)
\linethickness{0.3mm}
\put(3.2,2.6){\line(1,0){3.2}}
\put(0.9,-.7){\line(1,0){3.2}}
\end{picture}
|
\right),
\end{array}
\end{equation}
\end{widetext}
where the sums run over the 4-site and 6-site loops with all possible 
orientations. Although the repulsion is a higher order process, we have included
a repulsion term in the Hamiltonian, and we will treat its amplitude
$V$ as a free parameter to be able to make contact with the Rokhsar-Kivelson model
on the triangular lattice. The hopping amplitudes $t$ and $t^\prime$ are negative, and 
although the ratio $t^\prime/t$ is in principle fixed by their expression in the 
perturbative expansion, we will also treat it as a free parameter. 

Our central goal in this paper is to determine the nature of the ground state as
a function of $t^\prime/t$. With respect to what we already know
about QDMs, the main question is whether a competition between kinetic terms can also 
lead to a liquid phase. As we shall see, the answer to that question is positive, a 
liquid phase being present in a finite region of the phase diagram in the 
$t^\prime{-}V$ plane.

The paper is organized as follows. In Section~\ref{method}, we briefly review the basic 
preliminaries used in the rest of the paper. The results obtained with exact 
diagonalizations are presented in Section~\ref{resed}, those obtained with quantum 
Monte Carlo in Section~\ref{resqmc}, and the conclusions in Section~\ref{concl}. 
The perturbation calculation which has motivated the investigation of this QDM is finally
presented in Appendix~\ref{derivation}.

\section{The method}\label{method}

In this section, we present a brief introduction to the numerical methods, to the clusters used
in the analysis, and to the physical concepts underlying the determination of the phase diagram.
More details can be found in Ref.~\onlinecite{ralko}.

\begin{figure}
\includegraphics[width=0.45\textwidth]{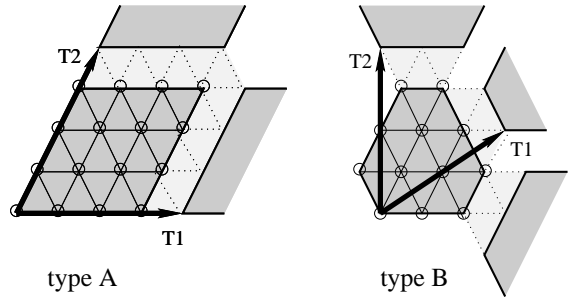}
\caption{\label{fig:cluster}
Example of type-A (left) and type-B (right) clusters with 16 and 12 sites, respectively.}
\end{figure}

Let us first discuss the shape of the finite-size clusters.
In general, a finite cluster is defined by two vectors ${\bf T}_1$ and ${\bf T}_2$ and,
in order to have the symmetries for rotations by $2\pi/3$, they have to satisfy:~\cite{bernu}
\begin{eqnarray}
{\bf{T}}_{1} &=& l {\bf{u}}_{1} + m {\bf{u}}_{2}  \nonumber \\
{\bf{T}}_{2} &=& -m {\bf{u}}_{1} + (l+m) {\bf{u}}_{2} \nonumber ,
\end{eqnarray}
where $l$ and $m$ are integers and ${\bf u}_1=(1,0)$ and ${\bf u}_2=(1/2,\sqrt{3}/2)$ are
the unitary vectors defining the triangular lattice. The number of sites in the cluster
is $N= l^2+m^2+lm$. In order to have also the axial symmetry, and therefore all the 
symmetries of the infinite lattice, we must take either $lm=0$ or $l=m$. The first 
possibility corresponds to type-A clusters (with the notation of Ref.~\onlinecite{ralko}),
with $N=l^2$ sites; the second one gives rise to type-B clusters, with
$N=3 \times l \times l$ (for examples of both cases, see Fig.~\ref{fig:cluster}).
Since for $t^\prime/t=0$ and $V/t=0$ the ground state belongs to a crystalline phase with
a 12-site unit cell,~\cite{moessner,ralko} we will restrict in this work
to clusters with a number of sites multiple of 12, in order not to frustrate 
this order. To limit the finite size effects related to the geometry of the 
clusters, we will concentrate on type-B clusters, which are always compatible with this 
order. Note that the 6-dimer loop kinetic term does not introduce further restrictions 
since all it requires is to be able to accommodate 6-site unit cells.

A very important concept in the QDM is the existence of topological sectors.
Indeed, in the triangular lattice, the Hilbert space is split into four disconnected 
topological sectors on a torus defined by the parity of the number of dimers cutting two 
lines that go around the two axes of the torus and denoted by $(p,q)$ with $p,q=0$ 
(respectively $1$) if this number is even (respectively odd). One can convince oneself, 
by a direct inspection of the effect of the 4-site and 6-site terms, that these numbers 
are conserved quantities under the action of the Hamiltonian of Eq.~(\ref{hamilt}). 
More generally, this is a consequence of the fact that the topological sectors are not 
coupled by any {\it local} perturbation. These topological sectors are extremely useful 
to distinguish between valence bond solids and spin liquids. Indeed, valence bond solids 
are only consistent with some topological sectors, whereas RVB spin-liquid phases are 
characterized by topological degeneracy. Therefore, the main goal will be to investigate 
whether, in the thermodynamic limit, the topological sectors are degenerate or not. 
In that respect, it is useful to remember that the $(0,1)$ and $(1,0)$ sectors are always 
degenerate with either $(0,0)$ or $(1,1)$ (depending on the cluster geometry) since they 
contain the same configurations rotated by an angle $\pi/3$.~\cite{ralko}
One can thus, without any loss of generality, restrict oneself to the analysis of the 
$(0,0)$ and $(1,1)$ sectors.
Therefore, we define the absolute value of the topological gap as:
\begin{equation}\label{topogap}
\Delta E = |E_{00} - E_{11}|,
\end{equation}
where $E_{00}$ and $E_{11}$ are the total ground-state energies for the topological sectors 
with $p=q=0$ and $p=q=1$, respectively. This gap is expected to scale to zero with the cluster
size in the RVB phase.

Finally, in order to detect a possible dimer order, we also consider the static dimer-dimer 
correlations
\begin{equation}\label{dimercorr}
D^{i,j}(r-r^\prime) = \langle D^i(r) D^j(r^\prime) \rangle,
\end{equation}
where $D^i(r)$ is the dimer operator defined as follows: It is a diagonal operator in the configuration
space that equals $1$ if there is a dimer from the site $r$ to the site $r+a_i$, with 
$a_1=(1,0)$, $a_2=(1/2,\sqrt{3}/2)$, or $a_3=(-1/2,\sqrt{3}/2)$ and vanishes otherwise.

The method used is the same one as that used by Ralko and collaborators~\cite{ralko} to determine 
the phase diagram of the Rokhsar-Kivelson QDM on the triangular lattice and our investigations are 
Lanczos diagonalizations and Green's function Monte Carlo (GFMC) simulations.
In particular, the GFMC is a zero-temperature stochastic technique based on the power method: 
Starting from a given wave function and by applying powers of the Hamiltonian, the ground state is 
statistically sampled to extract its energy and equal-time correlation functions.
In principle, as in other Monte Carlo algorithms, in order to reduce the statistical fluctuations, 
it is very useful to consider the importance sampling, through the definition of a suitable 
{\it guiding function}. 
Unfortunately, when dealing with dimer models, it is very hard to implement an accurate and, 
at the same time, efficient guiding function for the crystalline phases. 
This problem is particularly relevant when the 6-site term becomes dominant. In this case, our 
simulations suffer from wild statistical fluctuations, deteriorating the convergence of the GFMC. 
As a consequence, we are not able to reach the largest available size, $432$-site cluster, 
for all parameters $t^\prime/t$ and $V/t$. 
By contrast, given the simple form of the spin-liquid ground state (that reduces to a 
superposition of all the configurations with the same weight at the Rokshar-Kivelson point), 
in the disordered region we can use the guiding function with all equal weights
and obtain very small fluctuations (no fluctuation at the Rokshar-Kivelson point) 
and, therefore, excellent results with zero computational effort. 
Combining these facts, the loss of the GFMC convergence can be interpreted as a signal for the 
appearance of a crystalline phase. Of course, this is not a quantitative criterion, but, as 
it will be shown in the following, it gives reasonable insight into the emergence of a dimer order.

\begin{figure}
\includegraphics[width=0.40\textwidth]{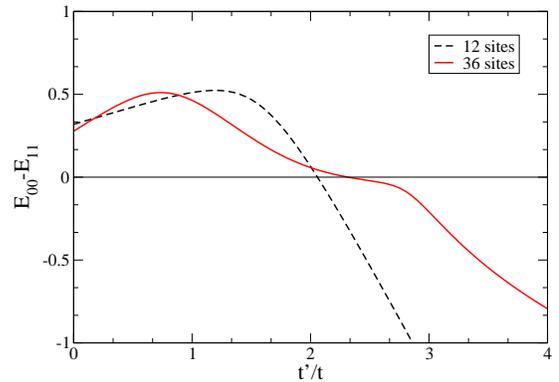}
\caption{\label{fig:ED_12_36}
(Color online) Difference between $E_{00}$ and $E_{11}$, the total ground-state energies of the 
topological sectors with $p=q=0$ and $p=q=1$, respectively. The results are found by exact 
diagonalizations for clusters with 12 and 36 sites.}
\end{figure}

\begin{figure}
\includegraphics[width=0.50\textwidth]{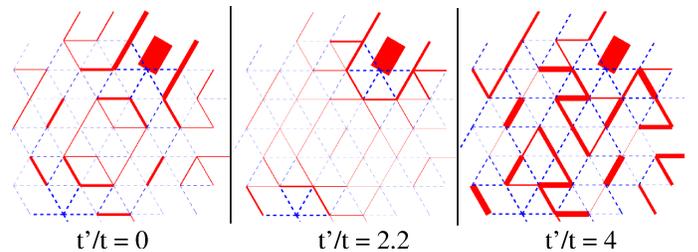}
\caption{\label{fig:correl}
(Color online) Dimer-dimer correlations on the 36-site cluster. The dimer of reference is 
the thickest one in the up-right corner. The thicker the line, the farther
the value of the correlation from the uniform distribution equal value $1/36$.
Solid lines are used when the correlation is higher than $1/36$, and dashed lines when 
it is lower.}
\end{figure}

Finally it should be mentioned that, since the different topological
sectors are completely decoupled (each dimer configuration belonging to one and only one of them) 
and cannot be connected by the terms contained in the Hamiltonian, within the GFMC it is possible to
work in a given topological sector, making it possible to extract the ground-state properties of 
each of them.
 
\section{Exact Diagonalizations}\label{resed}

To get a first idea of the properties of the model, we start with the results
we have obtained with exact diagonalizations of finite clusters for the model of 
Eq.~(\ref{hamilt}) with $V/t=0$.
Let us first begin with the ground-state energy for the 12- and 36-site clusters for both
the topological sectors $(0,0)$ and $(1,1)$ (see Fig.~\ref{fig:ED_12_36}). 
Note that the 12-site cluster is of type B, whereas the 36-site one is of type A. 
We have that, for both sizes, a level crossing occurs for $t^\prime/t \sim 2$. 
Below that value, the ground state is in $(1,1)$ topological sector, in agreement with 
the earlier results of Ref.~\onlinecite{ralko} for $V/t=0$.
The main difference between the 12-site and the 36-site clusters is that, for the 36-site 
cluster, the topological ground-state energies stay very close in a large parameter range 
for $t^\prime/t \sim 2$. This could suggest that, upon increasing the size, this level 
crossing might evolve into a phase where these energies are rigorously degenerate, giving
rise to a liquid phase without any crystalline order.

In order to give an idea of the various phases, we report in Fig.~\ref{fig:correl} the 
dimer-dimer correlations of Eq.~(\ref{dimercorr}) for the 36-site cluster below, at, and above 
the level crossing, which takes place at $t^\prime/t=2.2$ for this size. 
For small $t^\prime/t$, the correlations show a pattern similar to that of the intermediate 
$\sqrt{12} \times \sqrt{12}$ phase of the standard QDM model, already shown in 
Ref.~\onlinecite{ralko}. It should be stressed that, since in these calculations the Hamiltonian 
has the translational symmetry, the 12-site unit cell is not directly visible from 
Fig.~\ref{fig:correl}, and in order to have a clearer evidence one should break the symmetry by hand.
Nonetheless, as it has been shown in Ref.~\onlinecite{ralko}, these results are in perfect 
agreement with the existence of a $\sqrt{12} \times \sqrt{12}$ phase with a crystalline ground state
in the thermodynamic limit.
When $t^\prime/t$ is large, another pattern arises, which has never been observed 
in the standard QDM, and which presents a kind of 6-site triangle ordering. In this case, 
the dominant kinetic term involving 6 sites [see the second term of the Hamiltonian~(\ref{hamilt})]
induces a dimer pattern with the same symmetry, possibly inducing a
ground state with a 6-site unit cell in the thermodynamic limit. Also in this case the 
translational symmetry of the ground state partially masks the existence of a regular dimer pattern.
Unfortunately, we will not be able to confirm this prediction since, as stated before, the GFMC 
algorithm has serious problems of convergence inside this phase and for large clusters.

In any case, for intermediate values of $t^\prime/t$, the correlations decay 
very rapidly with the distance and are close to those obtained in the liquid phase of the
standard QDM with $V/t \lesssim 1$ and $t^\prime/t=0$.~\cite{ralko}
This fact gives another evidence of the possible existence of an RVB phase between two 
ordered phases in the model with competing kinetic terms and without the dimer repulsion. 
Of course by considering the exact diagonalization results only it is impossible to give 
definite statements on the stabilization of this liquid phase and, therefore, in the 
following section, we will consider a more systematic study of the topological gap, 
in order to unveil the existence of a wide disordered region that develops from the 
Rokshar-Kivelson point $V/t=1$ of the standard QDM and survives up to $V/t=0$ and 
finite $t^\prime/t$.

\section{Green's Function Monte Carlo}\label{resqmc}

In this section, we use the GFMC method to extend the results of the
previous section to larger clusters, with up to $432$ sites, and to map out the
phase diagram in the $t^\prime{-}V$ plane. 

\subsection{The case of $V/t=0$}

\begin{figure}
\includegraphics[width=0.45\textwidth]{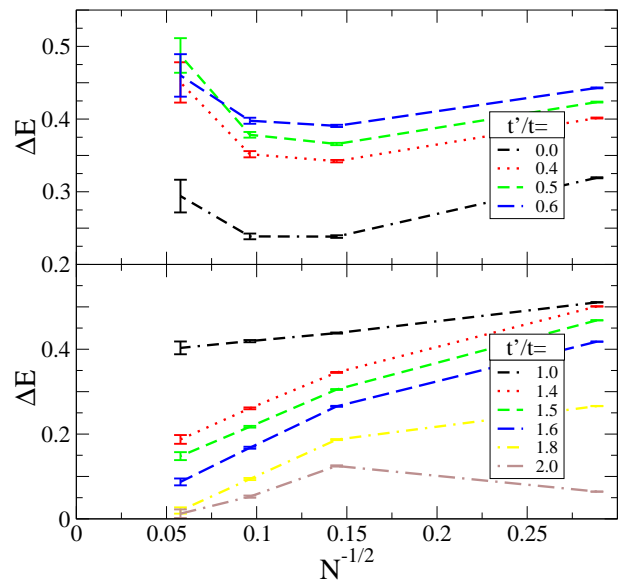}
\caption{\label{fig:topo_gap1}
(Color online) Topological gaps for $V/t=0$ as a function of $1/\sqrt{N}$, where $N$ is the 
number of sites and for different values of $t^\prime/t$. Upper panel: Small values of 
$t^\prime/t$, where the gap opens for large clusters. Lower panel: Larger values of 
$t^\prime/t$. For $t^\prime/t \gtrsim 1.6 $, the finite-size gap closes upon increasing the 
cluster size, signaling a liquid phase.}
\end{figure}

Let us first describe the results we have obtained for $V/t=0$ and consider the behavior of the
topological gap given in Eq.~(\ref{topogap}).
In Fig.~\ref{fig:topo_gap1}, for clarity, we divided the results in two sets for
$0 \leq t^\prime/t \leq 0.6$ and $t^\prime/t \geq 1$.
The first remarkable feature is that, for most parameters, the gap decreases between
12 and 48 sites, regardless of its behavior for larger sizes. Therefore, the possibility to 
study much larger clusters is crucial for this analysis. Indeed, there is a clear change 
of behavior for $t^\prime/t \sim 1.6$, a value beyond which our extrapolations give solid 
evidences in favor of a vanishing gap in the thermodynamic limit.
On the other hand, for smaller $t^\prime/t$ ratios, we have a clear evidence that $\Delta E$
increases for large sizes. Therefore, we come to the important conclusion that, 
the crystalline $\sqrt{12} \times \sqrt{12}$ phase is destroyed by increasing the 
amplitude of the 6-site term and the system is eventually driven into a liquid RVB phase.

\begin{figure}
\includegraphics[width=0.50\textwidth]{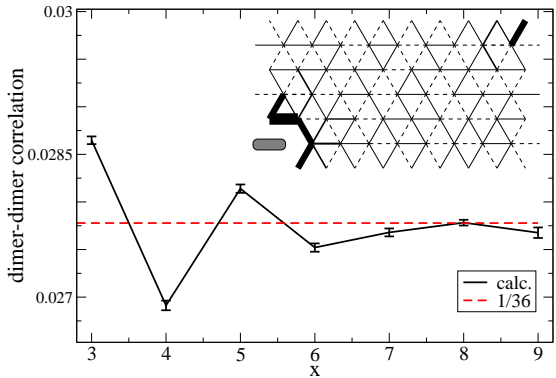}
\caption{\label{fig:parall_cs} 
(Color online) Dimer-dimer correlation function for a 108-site cluster along a 
horizontal line as a function of the distance for $t^\prime/t=2$ and $V/t=0$. The dashed 
line corresponds to $1/36$, the value in the absence of correlations.}
\end{figure}

To have a further confirmation of the existence of this disordered phase, we have calculated 
the dimer-dimer correlations. The results obtained on the 108-site cluster for 
$t^\prime/t=2$ are reported in Fig.~\ref{fig:parall_cs}, where the correlation functions
for parallel dimers along the same row are plotted as a function of the distance. 
Given the small number of clusters available, a precise size scaling of the order parameter 
is not possible and also a meaningful estimation of the correlation length is very hard.
Nevertheless, the behavior is definitely consistent with an exponential decay and the 
uncorrelated value of $1/36$ is approached very rapidly, as expected in a liquid phase 
without any crystalline order. 

Unfortunately, the larger $t^\prime/t$ region is numerically far more difficult to access. 
Indeed, as stated above, although the GFMC is in principle numerically exact, we have not 
been able to find an efficient guiding function to perform the importance sampling and
wild statistical fluctuations prevent us to reach a safe convergence for large clusters.
In practice, we have access to clusters up to 108 sites, that are still too small to 
predict the thermodynamic behavior. For instance, for $V/t=0$, the convergence stops
before one can observe any increase of the topological gap, and the criterion used
for the phase transition on the other side of the RVB phase cannot be used any more.
However, the lack of convergence is a clear sign that one enters a new (crystalline) phase.
So, if the change of behavior of the topological gap with the size cannot be observed, 
we take as a definition of the boundary for the phase transition the parameters for which 
the convergence is not good any more. 
This is expected to be semi-quantitative, and indeed, as we shall see in the next section 
when studying the full diagram, the points obtained with this criterion agree reasonably 
well with those obtained with the reopening of the topological gap.

In summary, although the region where the 6-site term dominates over the usual 4-site dimer
flip is not accessible by using GFMC, based on our numerical results for small $t^\prime/t$,
we can safely argue that the crystalline $\sqrt{12} \times \sqrt{12}$ phase is destabilized 
by increasing the 6-site kinetic term, leading to a true disordered ground state with
topological degeneracy.
 
\subsection{Phase diagram in the $t^\prime{-}V$ plane}

\begin{figure}
\includegraphics[width=0.45\textwidth]{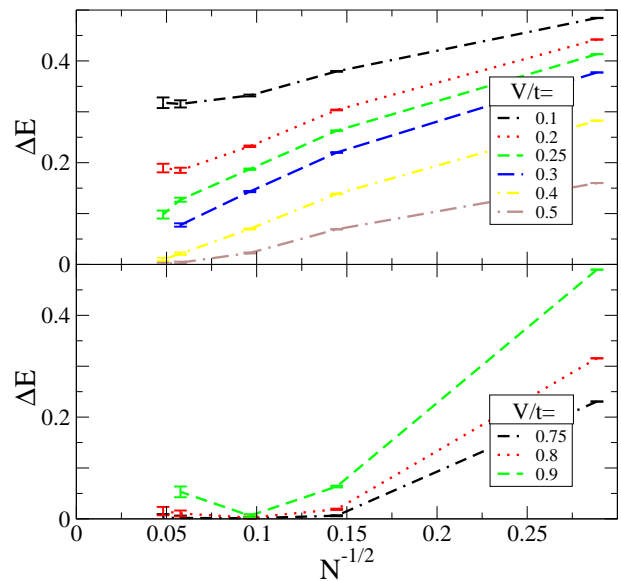}
\caption{\label{fig:topo_gap2} 
(Color online) Topological gap for $t^\prime/t=1$ and for various $V/t$ as a function of 
$1/\sqrt{N}$, where $N$ is the number of sites. Small and large values of $V/t$ 
have been shown separately in the upper and lower panels for clarity.}
\end{figure}

In this section we prove that the RVB phase found in the previous paragraph for $V/t=0$ is
connected to the one obtained for the standard QDM, i.e., close to the Rokshar-Kivelson
point and $t^\prime/t=0$. In order to do that, we have investigated a generalization of 
the previous model that also includes a repulsion $V$ between dimers facing each other, see
Eq.~(\ref{hamilt}). For $t^\prime/t=0$, this model reduces to the standard QDM, which has 
an RVB phase for $0.75 \lesssim V/t \lesssim 1$.~\cite{moessner,ralko}
To map out the complete phase diagram of this model, we have done the same analysis of the
previous paragraph for different values of $V/t$ between 0 and 1. As an example, we show in
Fig.~\ref{fig:topo_gap2} the finite-size scaling of the topological gap for $t^\prime/t=1$ and 
several values of $V/t$. The global behavior is the same as before and we have clear
evidence that the topological gap present at $V/t=0$ persists up to $V/t \sim 0.25$, and that
it opens again for $V/t \sim 0.8$. In that case, as
in many other cases, it turned out to be possible to actually observe the opening
of the gap upon leaving again the RVB phase before the convergence problems were too strong.

\begin{figure}
\includegraphics[width=0.40\textwidth]{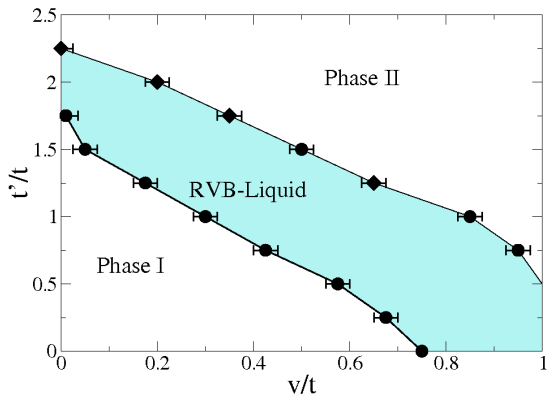}
\caption{\label{fig:phasediag} 
(Color online) Phase diagram in the $t^\prime{-}V$ plane. A wide disordered region
extends all the way from the standard QDM ($t^\prime/t=0$ axis) to the purely
kinetic QDM ($V/t=0$ axis). The description of the symbols is given in the text.}
\end{figure}

The resulting phase diagram is depicted in Fig.~\ref{fig:phasediag}, where different
symbols have been used depending on whether the boundary was determined from
the increase of the correlation length with the size or from the loss of convergence:
Solid circles when the closing of the topological gap was observable and
solid diamonds when the convergence of the GFMC was lost for large systems.
Interestingly enough, these different criteria build a relatively smooth line,
a good indication that it can be interpreted as a phase boundary.
Note that we did not perform simulations for $V/t>1$ where, for a vanishing $t^\prime/t$, 
a crystalline phase with staggered dimer order is stabilized. So we cannot exclude that 
the boundary extends beyond $V/t>1$ for small $t^\prime/t$. 
Remarkably, the RVB phase we found for $V/t=0$ is connected to the RVB phase
reported before for the standard QDM, i.e., $t^\prime/t=0$, the total RVB phase building 
up a large stripe that encompasses a significant portion of the phase diagram.  
We have also calculated static correlation function for several values of the
parameter, but they merely confirm the identification of the phases and are not
reported for brevity.

\section{Conclusions}\label{concl}

Coming back to our long-term motivation, namely to find an RVB phase in a 
realistic model of Mott insulators, this paper contains significant results of
two sorts. First of all, we have shown (see Appendix~\ref{derivation}) that, 
starting from the quasi-degenerate
mean-field ground state of a Kugel-Khomskii spin-orbital model, one can construct
a QDM with two remarkable properties: It describes a competition between
two kinetic terms of comparable magnitude, and all off-diagonal matrix elements in
the dimer basis are negative. This has allowed us to implement the GFMC
and to investigate the results of the competition between
these terms. It turns out that the competition between these kinetic terms leads to 
the disappearance of the $\sqrt{12}\times \sqrt{12}$ crystalline order when
$t^\prime/t \sim 1.6$. This transition is similar to the transition into the RVB phase that 
happens in the standard QDM upon approaching the Rokhsar-Kivelson point. Indeed,
the two phases can be connected into a single RVB phase in the context of a generalized QDM.
As far as the numerical investigation of the model is concerned, the main open issue
is to pin down the nature of the phase that occurs when the 6-dimer kinetic term dominates.
Unfortunately, the GFMC suffers from severe statistical fluctuations whenever the guiding
function is not accurate, i.e., for clusters larger that 108 sites and large $t^\prime/t$.
Therefore, we cannot make any definite statements on the phase where the 6-site term 
dominates. 
Another interesting question is of course the nature of the quantum phase transitions
between these phases (continuous or first order). We are currently working on that 
rather subtle issue in the context of the standard QDM. 

The general features of the RVB phase are consistent with the phenomenology
of LiNiO$_2$, which exhibits neither orbital nor magnetic long-range order. A number of 
points deserve further investigation however. The precise form of the QDM does not 
seem to be an issue: 
The actual model that can be derived along the Rokhsar-Kivelson lines has more terms 
(see Appendix~\ref{derivation}), but preliminary results show that the RVB phase is present 
in that model as well. 
The fact that the RVB phase does not contain the point $t^\prime/t=1.34$ derived in the 
Appendix~\ref{derivation} is not really an issue either. 
First of all, this ratio was determined for vanishing Hund's 
rule coupling and one vanishing hopping integral, and its precise value in that case should 
at best be taken as
an indication of its order of magnitude in the actual system. Besides, the other 6-site terms 
pull the RVB region down to smaller values of the relative ratio of the 6-dimer term
to the 4-dimer term. What would deserve more
attention is the validity of the 
expansion that leads to the effective QDM. The small parameter of the expansion is not 
that small ($\alpha=1/\sqrt{2}$, see Appendix~\ref{derivation}), and it would
be very useful to better understand to which extent such an expansion can be controlled.
Nevertheless, the present results strongly suggest that the presence of an RVB liquid phase 
between competing ordered phases is a generic feature of QDM, and that to identify such a 
phase in a realistic Mott insulator via an effective QDM is very promising.

We acknowledge useful discussions with P. Fazekas, M. Ferrero, and K. Penc. This work was 
supported by the Swiss National Fund and by MaNEP. F.B. is supported by CNR-INFM and MIUR
(COFIN 2005).

\appendix

\section{Derivation of the model}\label{derivation}

\begin{figure}
\includegraphics[width=0.15\textwidth]{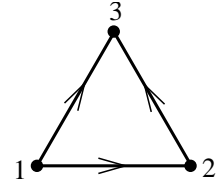}
\caption{\label{tri} 
Sign convention for a triangle}
\end{figure}

LiNiO$_2$ is a layered compound in which the Ni$^{3+}$ ions are in a low spin S=1/2 state
with a two-fold orbital degeneracy. A fairly general description of this system is given by
a Kugel-Khomskii Hamiltonian defined in terms of two hopping integrals $t_h$ 
and $t^\prime_h$, the on-site Coulomb repulsion $\tilde U$ and the Hund's coupling
$J$ which, on a given bond, takes the form~\cite{vernay}
\begin{widetext}
\begin{eqnarray}
\mathcal{H}_{ij} &=& 
  -\frac{2}{\tilde U\!+\!J} \left[
    2 t_h t^\prime_h {\bf T}_i {\bf T}_j
  - 4 t_h t^\prime_h T^y_i T^y_j 
    + (t_h-t^\prime_h)^2 ({\bf n}_{ij}^z{\bf T}_i) ({\bf n}_{ij}^z{\bf T}_j)\right.\nonumber\\ 
&&\left.  + \frac{1}{2} (t_h^2-{t^\prime_h}^2) \left( {\bf n}_{ij}^z{\bf T}_i
+ {\bf n}_{ij}^z{\bf T}_j \right) +  \frac{1}{4}(t_h^2+ {t^\prime_h}^2)  
    \right] \mathcal{P}_{ij}^{S=0} 
\nonumber\\
 && -\frac{2}{\tilde U -J} \left[
    4 t_h t^\prime_h T^y_i T^y_j 
  + \frac{1}{2}(t_h^2+ {t^\prime_h}^2)  
  + \frac{1}{2} (t_h^2-{t^\prime_h}^2) \left(
  {\bf n}_{ij}^z{\bf T}_i + {\bf n}_{ij}^z {\bf T}_j 
\right)
\right] \mathcal{P}_{ij}^{S=0} 
\nonumber\\
 && -\frac{2}{\tilde U\!-\!3J} \left[  - 2 t_h t^\prime_h {\bf T}_i {\bf T}_j
   - 
   (t_h-t^\prime_h)^2 ({\bf n}_{ij}^z{\bf T}_i) ({\bf n}_{ij}^z{\bf T}_j)+
\frac{1}{4}(t_h^2+{t^\prime_h}^2)
    \right] \mathcal{P}_{ij}^{S=1} 
\label{eq:effham3}
\end{eqnarray}
\end{widetext}
with the usual definitions for the projectors on the singlet and triplet states
of a pair of spins:
\begin{equation}
\mathcal{P}_{ij}^{S=0} = \frac{1}{4} - {\bf S}_i{\bf S}_j 
\quad\mbox{and}\quad
\mathcal{P}_{ij}^{S=1} = {\bf S}_i{\bf S}_j+\frac{3}{4},
\end{equation}
The vector ${\bf n}_{ij}^z$ depends on the type of bond. With the
convention of Fig.~\ref{tri}, they are given by:
\begin{equation}
\begin{array}{c}
{\bf n}_{12}^z=(0,0,1)\\
{\bf n}_{13}^z=(\frac{\sqrt{3}}{2},0,-\frac{1}{2})\\
{\bf n}_{23}^z=(-\frac{\sqrt{3}}{2},0,-\frac{1}{2}).
\end{array}
\end{equation}
The operators ${\bf T}_i$ are pseudo-spin operators acting on the orbitals.

On a given bond (see Fig.~\ref{tri}), 
a dimer is defined by the following wave function:
\begin{equation}
|\Phi_{ij}\rangle=|{\phi}_{ij}^\sigma\rangle\otimes|{\phi}_{ij}^\tau\rangle,
\end{equation}
where $|\phi_{ij}^\sigma\rangle$ and $|\phi_{ij}^\tau\rangle$ are 
respectively the spin and orbital components.
The spin component is the usual singlet given by:
\begin{equation}
|\phi_{ij}^\sigma\rangle=\alpha \left(|\uparrow_i\downarrow_j\rangle-
|\downarrow_i\uparrow_j\rangle\right),
\end{equation}
regardless of the orientation of
the bond. We have denoted by $\alpha$ the normalization coefficient, whose explicit value is
of course given by $1/\sqrt{2}$, to be able to keep track of the order in $\alpha$ of various 
overlaps. The orbital part depends on the bond and is given by:
\begin{eqnarray}
|\phi_{12}^\tau\rangle&=&|a_1\rangle|a_2\rangle\\
|\phi_{13}^\tau\rangle&=&|-\frac{1}{2}a_1-\frac{\sqrt{3}}{2}b_1\rangle
|-\frac{1}{2}a_3-\frac{\sqrt{3}}{2}b_3\rangle\\
|\phi_{23}^\tau\rangle&=&|-\frac{1}{2}a_2+\frac{\sqrt{3}}{2}b_2\rangle
|-\frac{1}{2}a_3+\frac{\sqrt{3}}{2}b_3\rangle
\end{eqnarray}
with the convention of Fig.~\ref{tri}, and with $|a\rangle=|d_{3z^2-r^2}\rangle$ 
and $|b\rangle=|d_{x^2-y^2}\rangle$.
We also use the convention that the wave function has a plus sign if the two sites
are in the order defined by the arrows of Fig.~\ref{tri}, and a minus
sign otherwise.

\begin{figure}
\includegraphics[width=0.20\textwidth]{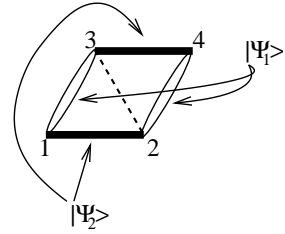}
\caption{\label{b4} 
A 4-dimer loop}
\end{figure}

\begin{figure}
\includegraphics[width=0.40\textwidth]{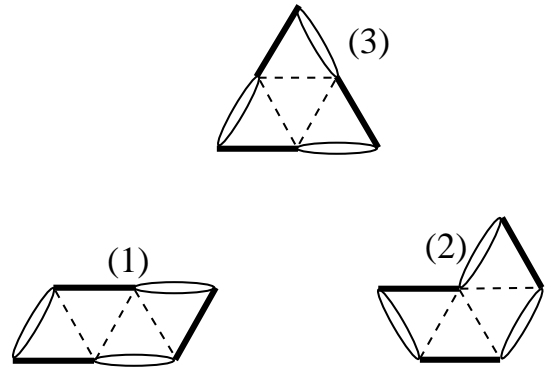}
\caption{\label{b6} 
The 3 types of 6-dimer loops}
\end{figure}

With these definitions, the matrix element of the Hamiltonian on each bond is given by:
\begin{equation}
\langle \Phi_{ij}|{\cal H}_{ij}|\Phi_{ij}\rangle=\epsilon\ 2\alpha^2=
\epsilon\ \langle \Phi_{ij}|\Phi_{ij}\rangle,
\end{equation}
where $\epsilon$ is defined by:
\begin{equation}
\epsilon=\frac{-4\tilde Ut_h^2}{\tilde U^2-J^2}<0
\end{equation}

According to the mean-field analysis of Ref.~\onlinecite{vernay}, the wave functions obtained as
tensor products of these wave functions on all dimer coverings on the triangular lattice
should constitute a good variational basis if $t^\prime_h\ll t_h$ and $J\ll \tilde U$. 
In the following, we consider for simplicity the case 
$t^\prime_h=0$ and $J=0$.

Following Rokhsar and Kivelson,~\cite{rokhsar}
the idea is now to perform a unitary transformation to derive the effective QDM.
If one defines the overlap matrix by
$S_{mn}=\langle\Psi_m|\Psi_n\rangle$, where $|\Psi_m\rangle$ and $|\Psi_n\rangle$
are dimer coverings, the states defined by
\begin{equation}
| m \rangle = \sum_n \left(S^{-\frac{1}{2}}\right)_{m,n} | \Psi_n \rangle
\end{equation}
constitute an orthonormal basis and the matrix elements of the Hamiltonian in
this basis are given by:
\begin{equation}\label{hamilteff}
{H}^{\textrm{eff}}_{mn} \equiv \langle m| {\cal H} |n \rangle =\sum_{kl}\left(S^{-\frac{1}{2}}\right)_{mk}
\langle\Psi_k|{\cal H}|\Psi_{l}\rangle
\left(S^{-\frac{1}{2}}\right)_{ln}.
\end{equation}
The inverse of the square root of the overlap matrix cannot be calculated exactly, but this 
can be done approximately in the context of an expansion in powers of $\alpha$. Indeed,
the overlap matrix can be expanded as: 
\begin{equation}\label{overlap}
S= I + {2} A \alpha^4+ {2} B \alpha^6+O[\alpha^8],
\end{equation} 
which leads to:
\begin{equation}
\left(S\right)^{-\frac{1}{2}}= I - A\alpha^4 - B\alpha^6+O[\alpha^8].
\end{equation} 
In these expressions, the matrices $A$ and $B$ only have non vanishing matrix
elements,  equal to $1$,
between configurations that are the same except on  
a 4-dimer loop (see Fig.~\ref{b4}) or on one of the three types of 6-dimer loop (see 
Fig.~\ref{b6}), respectively. 

\begin{table}
\caption{\label{tablo} 
Overlap and matrix element for two configurations
differing by a loop-4, a loop-6 of type (1), (2) and (3).}
\begin{tabular}{|*{5}{c|}}
\hline
 & loop-4 & loop-6 (1)& loop-6 (2) & loop-6 (3) \\
\hline
$\langle m | {\cal H} | n \rangle$ & $- \frac{66 \epsilon \alpha^4}{16^2}$ &
$- \frac{103 \epsilon \alpha^6}{16^2}$ & $- \frac{24 \epsilon \alpha^6}{16^2}$ &
$- \frac{177 \epsilon \alpha^6}{16^2}$ \\
 & & & & \\
& $\simeq -0.0644 \epsilon$ & $\simeq -0.0503 \epsilon$ & $\simeq -0.0012 \epsilon
$ & $\simeq -0.0864 \epsilon$ \\
\hline
\end{tabular}
\end{table}

Similarly, the Hamiltonian matrix $\tilde{\cal H}$ defined by 
$\tilde{\cal H}_{mn}=\langle\Psi_m|{\cal H}|\Psi_n\rangle - \epsilon N_d
\delta_{mn}$, where $N_d$ is the number of dimers,
has an expansion in powers of $\alpha$ that reads:
\begin{equation}\label{expansionH}
\tilde{\cal H}= C\alpha^4+ D\alpha^6+O[\alpha^8],
\end{equation}
where the matrices $C$ and $D$ have non vanishing 
matrix elements under the same conditions as matrices $A$ and $B$.

Since all these expansions start with $\alpha^4$, it is clear that the first contribution to the 
diagonal part of ${ H^{\textrm{eff}}}$ will be of order $\alpha^8$. So, 
to order $\alpha^6$, the effective Hamiltonian will only have off-diagonal
matrix elements. Moreover, to this order, these matrix elements are simply given by 
$\langle m|\tilde{\cal H}|n\rangle=\langle\Psi_m|\tilde{\cal H}|\Psi_n\rangle$. They are tabulated 
in Table~\ref{tablo} for configurations that are the same except on  
a 4-dimer loop (Fig.~\ref{b4}) or on one of the three types of 6-dimer loop (Fig.~\ref{b6}).
All these matrix elements are negative. Among the 6-dimer loop terms, the matrix
element of type (3) is the largest, and its ratio to the 4-dimer loop term is 1.34.



\begin{thebibliography}{99}
\bibitem{rokhsar} D.S. Rokhsar and S.A. Kivelson, \prl {\bf 61}, 2376 (1988).
\bibitem{leung} P.W. Leung and K.C. Chiu, \prb {\bf 54}, 12938 (1996).
\bibitem{syljuasen} O.F. Sylju\aa sen, \prb {\bf 71}, 020401(R) (2005).
\bibitem{moessner} R. Moessner and S.L. Sondhi, \prl {\bf 86}, 1881 (2001).
\bibitem{ralko} A. Ralko, M. Ferrero, F. Becca, D. Ivanov, and F. Mila, \prb {\bf 71}, 
224109 (2005).
\bibitem{balents} L. Balents, M.P.A. Fisher, and S.M. Girvin, \prb {\bf 65}, 224412 (2002).
\bibitem{misguich} G. Misguich, D. Serban, and V. Pasquier, \prl {\bf 89}, 137202 (2002).
\bibitem{mila} F. Mila, \prl {\bf 81}, 2356 (1998).
\bibitem{zhitomirsky} M.E. Zhitomirsky, \prb { \bf 71}, 214413 (2005).
\bibitem{kugel} K.I. Kugel and D.I. Khomskii, Sov. Phys. Usp.  {\bf 25}, 232 (1982).
\bibitem{vernay} F. Vernay, K. Penc, P. Fazekas, and F. Mila, \prb {\bf 70}, 014428 (2004).
\bibitem{bernu} B. Bernu, P. Lecheminant, C. Lhuillier, L. Pierre, \prb {\bf 50}, 10048 (1994).
\bibitem{anderson} P.W. Anderson, Mater. Res. Bull. {\bf 8}, 153 (1973).
\bibitem{fazekas} P. Fazekas and P.W. Anderson: Philos. Mag. {\bf 30}, 423 (1974).
\bibitem{ioselevich} A. Ioselevich, D.A. Ivanov, and M.V. Feigel'man, \prb {\bf 66}, 
174405 (2002).
\bibitem{moessner2} R. Moessner and S.L. Sondhi, \prb {\bf 63}, 224401 (2001).
\end{thebibliography}
\end{document}